\documentclass[conference]{IEEEtran}
\IEEEoverridecommandlockouts
\usepackage{cite}
\usepackage{amsmath,amssymb,amsfonts}
\usepackage{algorithmic}
\usepackage{graphicx}
\usepackage{textcomp}
\usepackage{xcolor}
\usepackage{diagbox}
\def\BibTeX{{\rm B\kern-.05em{\sc i\kern-.025em b}\kern-.08em
    T\kern-.1667em\lower.7ex\hbox{E}\kern-.125emX}}
\usepackage{amsmath, amssymb, bm, cite, epsfig, psfrag}
\usepackage{graphicx}
\usepackage{soul} 
\usepackage[margin=0.7in]{geometry}
\usepackage{dblfloatfix}
\usepackage{array}
\usepackage{lipsum}
\usepackage{gensymb}
\newcolumntype{P}[1]{>{\centering\hspace{0pt}}p{#1}}
\newcolumntype{M}[1]{>{\centering\hspace{0pt}}m{#1}}
\newcolumntype{L}{>{\centering\arraybackslash}m{3cm}}
\usepackage[font=small]{caption}
\usepackage{epstopdf}	
\usepackage{longtable}
\usepackage{supertabular,booktabs}
\usepackage{bbm}
\usepackage{multirow}

\usepackage{subfigure}
\usepackage{etoolbox}
\usepackage{pbox}
\usepackage{fixltx2e}
\usepackage{tabu}	
\usepackage{filecontents}
\usepackage{enumerate}
\usepackage{textcomp}
\usepackage{float}
\usepackage{colortbl}
\usepackage{fancyhdr}

\usepackage{bm}
\newtoggle{conference}
\togglefalse{conference} 
\usepackage{cuted}
\fancypagestyle{firststyle}{
	\fancyhf{}
	\fancyhead[L]{S. Ju, O. Kanhere, Y. Xing and T. S. Rappaport, ``A Millimeter-Wave Channel Simulator NYUSIM with Spatial Consistency and Human Blockage,''  2019 \textit{IEEE Global Communications Conference (GLOBECOM)}, Hawaii, USA, Dec. 2019, pp. 1-6.}     

}

\begin{document}

\title{A Millimeter-Wave Channel Simulator NYUSIM with Spatial Consistency and Human Blockage}

\author{\IEEEauthorblockN{Shihao Ju, Ojas Kanhere, Yunchou Xing and Theodore S. Rappaport}
\IEEEauthorblockA{\textit{NYU Tandon School of Engineering, NYU WIRELESS, Brooklyn, NY, 11201}\\
\{shao, ojask, ychou, tsr\}@nyu.edu}
\thanks{This work is supported in part by the NYU WIRELESS Industrial Affiliates, and in part by the National Science Foundation under Grants: 1702967 and 1731290.}
}

\maketitle
\thispagestyle{firststyle}
\begin{abstract}
Accurate channel modeling and simulation are indispensable for millimeter-wave wideband communication systems that employ electrically-steerable and narrow beam antenna arrays. Three important channel modeling components, spatial consistency, human blockage, and outdoor-to-indoor penetration loss, were proposed in the 3rd Generation Partnership Project Release 14 for mmWave communication system design. This paper presents NYUSIM 2.0, an improved channel simulator which can simulate spatially consistent channel realizations based on the existing drop-based channel simulator NYUSIM 1.6.1. A geometry-based approach using multiple reflection surfaces is proposed to generate spatially correlated and time-variant channel coefficients. Using results from 73 GHz pedestrian measurements for human blockage, a four-state Markov model has been implemented in NYUSIM to simulate dynamic human blockage shadowing loss. To model the excess path loss due to penetration into buildings, a parabolic model for outdoor-to-indoor penetration loss has been adopted from the 5G Channel Modeling special interest group and implemented in NYUSIM 2.0. This paper demonstrates how these new modeling capabilities reproduce realistic data when implemented in Monte Carlo fashion using NYUSIM 2.0, making it a valuable measurement-based channel simulator for fifth-generation and beyond mmWave communication system design and evaluation. 

\end{abstract}
    

\section{Introduction}~\label{sec:intro}

The millimeter-wave (mmWave) spectrum is regarded as a promising band to support the unprecedented capacity demand due to the massive available bandwidth \cite{Rap13a}. The directional mmWave channel has vastly different channel statistics as compared to the semi-omnidirectional and sectored microwave channels \cite{Sun18a}. Accurate channel modeling for mmWaves is essential for the fifth-generation (5G) and beyond wireless communication system design and evaluation \cite{Sun18a}. Many promising applications will be enabled using mmWave and sub-Terahertz technologies such as wireless cognition, imaging, and precising positioning \cite{Rap19access}. 

The scattering environment is similar when a user terminal (UT) moves in a local area or when multiple UTs are closely spaced in a local area (e.g. within 10-15 m) \cite{Ju18a,Ju18b}. Further, the channel impulse responses (CIRs) of these locations in close proximity to each other should be highly correlated. A channel model with spatial consistency can generate correlated and time-variant channel coefficients along the UT trajectory \cite{Ju18a,Ju18b,Wang15a,Ademaj17a}. A channel simulator with spatial consistency can simulate consecutive angular power spectrums and power delay profiles (PDPs) in a continuous and realistic manner based on the UT trajectory within a local area. 

Human blockage becomes an important factor in radio signal strength for mmWave communication systems, but did not attract much attention in the microwave (sub-6 GHz) communications era \cite{Mac17b}. Owing to very short wavelengths (a few millimeters) and the use of directional antennas, mmWaves are easily blocked by humans and do not effectively diffract around human bodies or vehicles. It is important to take shadowing loss caused by humans and vehicles in account for accurate link budget analysis \cite{Mac19a}.

Outdoor-to-indoor (O2I) penetration loss becomes more prominent at mmWave frequencies as shown in measurements and models \cite{Rap13a,Xing19a,Haneda16a}. Many modern buildings are constructed by concrete and have infrared reflecting (IRR) glass, which induce a large penetration loss when a mmWave signal is transmitted from outdoor to indoor or vice versa \cite{Haneda16a}. Thus, accurate O2I penetration loss prediction is also critical for the design and deployment of future outdoor and indoor mmWave communication systems \cite{Andrew17a}.

The rest of the paper is organized as follows. Section \ref{sec:nyusim_old} reviews the channel model and simulator of the previous NYUSIM 1.6.1 \cite{Sun17b}. Section \ref{sec:nyusim_new} explains the extended NYUSIM channel model with spatial consistency, human blockage, and O2I penetration. Numerical results for model validation are also presented in Section \ref{sec:nyusim_new}. Section \ref{sec:gui} introduces the user interface, editable parameters, and sample output figures and data files of NYUSIM 2.0. Concluding remarks are provided in Section \ref{sec:conclusion}.

\section{Drop-based NYUSIM channel simulator}~\label{sec:nyusim_old}
NYUSIM is an open-source mmWave channel simulator \cite{Sun17b}, which can produce accurate omnidirectional and directional CIRs, PDPs, and \textcolor{black}{3-dimensional (3-D)} angular power spectrum. NYUSIM is developed based on extensive field measurements from 28 GHz to 140 GHz \cite{Rap13a, Rap15b}. A 3-D spatial statistical channel model forms the basis for, and is implemented in NYUSIM \cite{Samimi16a}, which characterizes temporal and angular properties of multipath components (MPCs). NYUSIM can operate over a wide range of carrier frequencies from 500 MHz to 100 GHz and support wide RF bandwidth up to 800 MHz. Different types of antenna arrays are also supported such as uniform linear array (ULA) and uniform rectangular array (URA). Some key features of the drop-based NYUSIM is summarized as follows. 

\paragraph{Drop-based} In a drop, the drop-based channel model generates a static and independent CIR at a particular transmitter-receiver (T-R) separation distance. However, there is no correlation among different drops. A set of identical and independent distributed (i.i.d) large-scale parameters (LSPs) and small-scale parameters (SSPs) is generated for each channel drop. The shortcoming of a drop-based channel model is that it generates independent channel coefficients for different distances, even if these distances are close together. 
	
\paragraph{Large-scale Path Loss Model } The close-in free space reference distance (CI) path loss model with a 1 m reference distance \cite{Rap15b,Sun16a,Mac15a}, and an extra attenuation term due to various atmospheric conditions, is employed in NYUSIM, which is given by \cite{Rap15b,Sun16a}:
\begin{equation}
	\label{eq:pathloss}
	\begin{split}
	\textup{PL}^{\textup{CI}}(f,d)[\textup{dB}]=&\textup{FSPL}(f,1 m)[\textup{dB}]+10n\log_{10}(d)\\
	&+\textup{AT}[dB]+\chi_\sigma,
	\end{split}
\end{equation}
where $d$ is the 3-D T-R separation distance in meter, and $n$ is the path loss exponent ($n$ = 2 for free space). $\chi_\sigma$ is the shadow fading (SF) modeled as a log-normal random variable with zero mean and $\sigma$ standard deviation in dB. $\textup{AT}$ is a total atmospheric absorption term \cite{Liebe93a}. $\textup{FSPL}(f,1 \textup{m})$ is the free space path loss in dB at a T-R separation distance of 1 m at the carrier frequency $f$ in GHz \cite{Sun17b}:
\begin{equation}
	\begin{split} 
	\textup{FSPL}(f,1 m)[\textup{dB}]&=20\log_{10}(\frac{4\pi f\times10^9}{c})\\
	&=32.4[\textup{dB}]+20\log_{10}(f),
	\end{split}
\end{equation}
where $c$ is the speed of light. 
	
\paragraph{Wideband Temporal/Spatial Clustering Algorithm} A time-cluster spatial-lobe (TCSL) approach is adopted in the NYUSIM channel model to characterize the temporal and spatial properties \cite{Samimi16a}, which is motivated from the field measurements \cite{Rap13a,Rap15b}. A time cluster is defined as a group of MPCs traveling close in time, but may arrive from different angular directions. A spatial lobe is defined as a group of MPCs coming from a similar direction, but may arrive in a long time window of several hundreds of nanoseconds. In other words, the temporal and spatial properties of MPCs are separately generated, different from joint time-angle distributions adopted in current 3GPP model \cite{3GPP.38.901}.

\vspace{-0.1cm}
\section{New Features Implmented in NYUSIM 2.0}~\label{sec:nyusim_new}
\vspace{-0.8cm}
\subsection{Spatial Consistency Procedure}
Spatial consistency indicates continuous and realistic channel evolution along the UT trajectory in a local area. To realize spatial consistency, spatially correlated large-scale parameters such as SF, line-of-sight (LOS)/non-LOS (NLOS) condition are generated, and time-variant small-scale parameters such as angles, power, delay, phase of each MPC are generated \cite{Ju18a,Ju18b}. The correlation distance of LSPs precisely characterizes the concept of ``a local area'', which also confines the length of a channel segment (where a channel segment is 10-15 m long) \cite{Ju18b}. In a channel segment, the channels are considered highly correlated and updated using spatial consistency procedure. A channel segment can be divided into several channel snapshots. The distance between two channel snapshots is the update distance (e.g. 1 m), which means the channel coefficients are updated for every 1 m increment along a traveled path. 
\vspace{-1em}
\begin{figure}[!htbp]
	\centering
	\setlength{\abovecaptionskip}{0cm}
	\setlength{\belowcaptionskip}{-0.4cm}
	\includegraphics[width=0.35\textwidth]{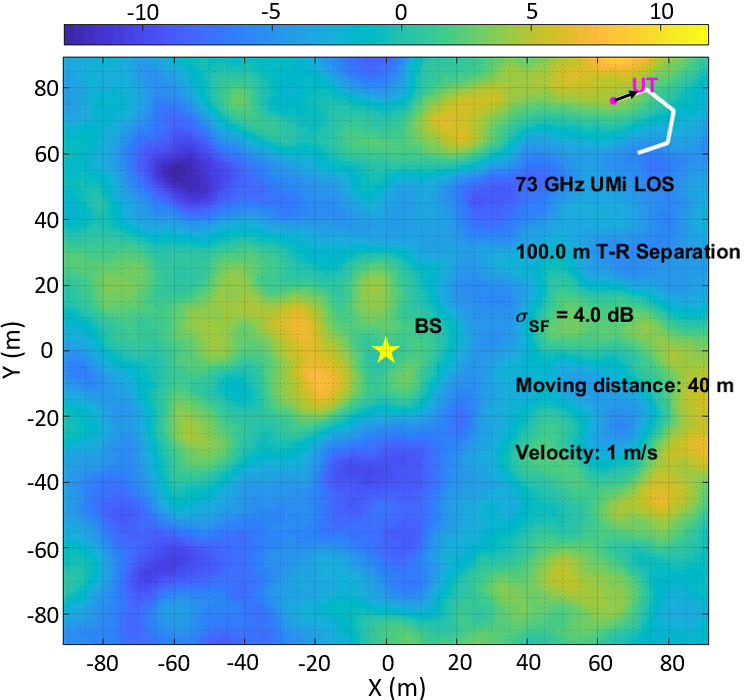}
	\caption{A map of spatially correlated SF with the BS and UT locations. The map of spatially correlated SF is generated by filtering a map of independent SF using an exponential function in (\ref{eq:expo_fun}). SF [dB]$\sim$N(0,4) in a UMi LOS scenario. T-R separation distance is 100 m. } \label{fig:sf_map}
\end{figure}
\vspace{-1em}
\begin{figure}[!htbp]
	\centering
	\setlength{\abovecaptionskip}{0.1cm}
	\setlength{\belowcaptionskip}{-0.4cm}
	\includegraphics[width=0.35\textwidth]{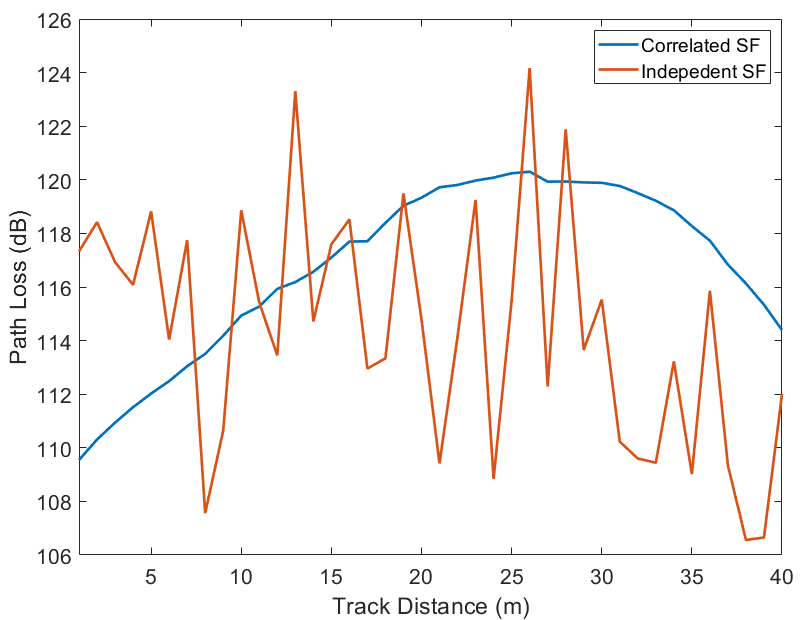}
	\caption{The UT moves in a partial hexagon track in a clockwise manner as shown in Fig. \ref{fig:sf_map}. The track distance is 40 m, the side length of the hexagon is 10 m. } \label{fig:path_loss}
\end{figure}
\subsubsection{Spatially Correlated Large-scale Parameters}
\label{sec:lsp}

LSPs defined in the 3GPP 38.901 model are delay spread, angular spread, Rician K factor, SF, and LOS/NLOS condition \cite{3GPP.38.901}. \textcolor{black}{The two LSPs explicitly used in the channel generation procedure in NYUSIM is SF and LOS/NLOS condition. Spatially correlated SF and LOS/NLOS condition values are generated in the same way. For SF,} a 2-dimensional (2-D) grid map is generated to contain values of spatially correlated SF in a simulated area. The granularity of the map is set to be 1 m, which means the distance between two neighboring grid points is 1 m. SF is modeled as a log-normal random variable with zero mean and $\sigma$ dB standard deviation as shown in (\ref{eq:pathloss}). The map of SF is initialized by assigning an i.i.d normal distributed random variable at each grid. \textcolor{black}{A 2-D exponential filter is applied to the map, which is given by \cite{Ju18a}:
\begin{equation}
\label{eq:expo_fun}
	h(p,q) = \exp(-\frac{\sqrt{p^2+q^2}}{d_{\textup{co}}}),
\end{equation} 
where $p$ and $q$ are coordinates with respect to the center of the filter. $\sqrt{(p^2+q^2)}$ represents the distance to the center of the filter. $d_{\textup{co}}$ is the correlation distance of SF. $L$ is the length of the filter, which is set as 8-fold $d_{co}$ since the correlation is negligible beyond 4-fold $d_{co}$. Applying this 2-D filtering, the correlated values in the map is calculated by:
\begin{equation}
\label{eq:2d_filter}
	M_c(i,j) = \sum_{p}\sum_{q}h(p,q)M(i-p+1,j-q+1)
\end{equation}
where $M_c$ is the correlated map and $M$ is the initialized independent map. $i$ and $j$ are the coordinates of grid points in the map.}

A map of spatially correlated SF over a 200 m x 200 m area is shown in Fig. \ref{fig:sf_map}. The correlation distance of SF in a UMi LOS scenario is set to be 10 m \cite{Ju18b}. The UT moved 40 m in a partial hexagon track, which is also illustrated in Fig. \ref{fig:sf_map}. SF varies from -10 dB to 10 dB in a continuous manner. Similar SF values are observed at closely spaced locations whereas independent values for close locations are always used in the drop-based model. The time-variant path loss is shown in Fig. \ref{fig:path_loss}. It can be seen that the path loss varies rapidly about 16 dB with independent SF values. However, the path loss varies smoothly with spatially correlated SF, which corresponds to the UT trajectory. The path loss increases first and then drops as the UT first moves away from the base station and then comes back. Spatially correlated SF supports a more realistic link budget analysis and cellular coverage prediction. The map of spatially correlated LOS/NLOS condition is also created by applying a 2-D exponential filter. 

\subsubsection{Time-variant Small-scale Parameters}

Small-scale parameters include angles, delay, phase, power of each MPC, which need to be realistically updated based on the UT trajectory every channel snapshot. A geometry-based approach is adopted in NYUSIM to update small-scale parameters:
\paragraph{Update of angles} 
Each MPC has four angles, azimuth angle of departure (AOD), \textcolor{black}{zenith} angle of departure (ZOD), azimuth angle of arrival (AOA), and \textcolor{black}{zenith} angle of arrival (ZOA). A LOS component can be simply updated using geometry based on the BS and UT locations and UT moving speed and direction. A simple linear update is given by \cite{Wang15a,Ju18a,3GPP.38.901}:
\begin{equation}
    \phi_{\textup{angle}}(t_k) = \phi_{\textup{angle}}(t_{k-1})+ S_{\textup{angle}}\cdot \Delta t,
\end{equation}
where ``angle'' can be AOD, ZOD, AOA, and ZOA. $\Delta t=t_k-t_{k-1}$. The update part $S_{\textup{angle}}$ for a LOS component can be given by \cite{3GPPTDOC1707267}:
\vspace{-0.5cm}
\begin{strip}
\begin{align}
	S_{\textup{AOD}} &= \frac{v_y\cos(\phi_{\textup{AOD}})-v_x\sin(\phi_{\textup{AOD}})}{r\sin(\theta_{\textup{ZOD}})}, S_{\textup{AOA}} = \frac{v_y\cos(\phi_{\textup{AOA}})-v_x\sin(\phi_{\textup{AOA}})}{r\sin(\theta_{\textup{ZOA}})},\\	
	S_{\textup{ZOD}} &= \frac{v_x\cos(\phi_{\textup{AOD}})\cos(\theta_{ZOD})+v_y\cos(\theta_{\textup{ZOD}})\sin(\phi_{\textup{AOD}})-v_z\sin(\theta_{\textup{ZOD}})}{r}, \\
	S_{\textup{ZOA}} &= \frac{v_x\cos(\phi_{\textup{AOA}})\cos(\theta_{ZOA})+v_y\cos(\theta_{\textup{ZOA}})\sin(\phi_{\textup{AOA}})-v_z\sin(\theta_{\textup{ZOA}})}{r},
\end{align}
\end{strip}
\noindent where $\phi_{\textup{AOD}}$, $\theta_{\textup{ZOD}}$, $\phi_{\textup{AOA}}$, $\theta_{\textup{ZOA}}$ are AOD, ZOD, AOA, and ZOA, respectively. $v_x$, $v_y$, $v_z$ are the projections of the velocity vector $\mathbf{v}$ in x, y, z-axis, respectively. 
\begin{figure}[!htbp]
	\centering
		\setlength{\abovecaptionskip}{0cm}
	\setlength{\belowcaptionskip}{-0.4cm}
	\includegraphics[width=0.35\textwidth]{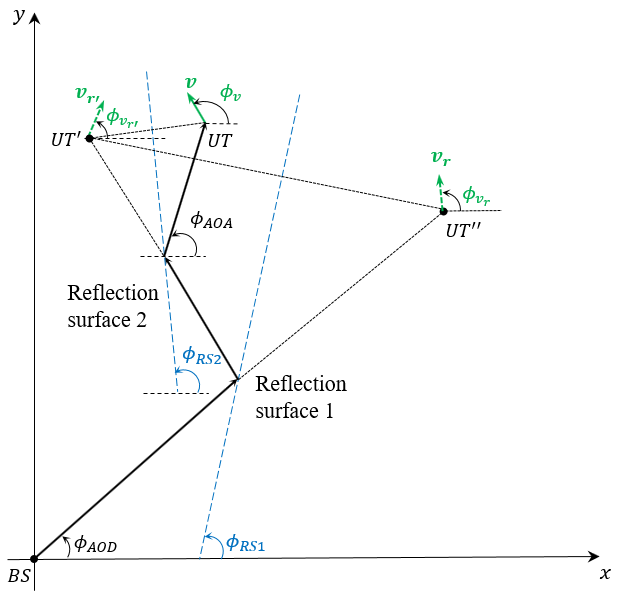}
	\caption{UT movement and change of AOD in the azimuth plane in NLOS scenarios for a multiple reflection case.} \label{fig:nlos_phi_2}
\end{figure}

For NLOS components, motivated by a simple case illustrated in \cite{Wang15a}, a multiple reflection surfaces (MRS) method is proposed to transform the $UT$ location in the NLOS scenario to the $UT'$ location in the LOS scenario from the BS. Then, NLOS components can be considered as virtual LOS components and updated using the same equations as LOS components. An illustration of a NLOS component that experienced one reflection can be found in \cite{Wang15a,Ju19b}. $UT'$ and $\phi_{vr}$ are the mirror images of $UT$ and the moving direction $\phi_v$. The relations in one reflection case are given as:
\begin{align}
\label{eq:reflect_odd}
\phi_{\textup{AOA}} &= 2\phi_{\textup{RS}}+\pi-\phi_{\textup{AOD}}\\
\phi_{v_r} &= 2\phi_{\textup{RS}}-\phi_v
\end{align}

Similarly, a NLOS component that experienced two reflections is shown in Fig. \ref{fig:nlos_phi_2}, and the angle relations in two reflections case are given as:
\begin{align}
\label{eq:reflect_even}
\phi_{\textup{AOA}} &= 2\phi_{RS2}-2\phi_{\textup{RS}}+\phi_{\textup{AOD}} = \Delta_{\textup{RS}}+\phi_{\textup{AOD}}\\
\phi_{vr} &= 2\phi_{RS2}-2\phi_{\textup{RS}}+\phi_v = \Delta_{\textup{RS}}+\phi_v
\end{align}
Further for a NLOS component that experienced $M$ reflections, the relations can be derived by iteration:
\begin{equation}
\label{eq:reflect_surface}
\begin{split}
\phi_{\textup{AOA}} &= (-1)^M\phi_{\textup{AOD}} + 2\sum_{i=1}^{M}(-1)^i\phi_{RSi}+M\pi \\
&= \Delta_{\textup{RS}}+ (-1)^M\phi_{\textup{AOD}} + M\pi\\
\end{split}
\end{equation}
\begin{equation}
	\phi_{vr} = \Delta_{\textup{RS}} + (-1)^M\phi_v
\end{equation}

As a conclusion, there are two cases depending on that a NLOS component experienced odd or even number of reflection surfaces. A random binary number $B$ can be used to represent two cases, when $B=-1$ represents even number of reflection surfaces, and $B=1$ represents odd number of reflection surfaces. 
\paragraph{Update of delays and phases}
The delay and phase of each MPC is updated simply based on the path length change using Law of cosines. For a LOS component, the path length is the T-R separation distance. For NLOS components, the path lengths are obtained by multiplying the initialized absolute time delay with the speed of light. 
\paragraph{Update of powers}
The power of each MPC is updated by redistributing cluster powers and MPC powers in each cluster following the same way that these powers were initialized using the cluster time excess delay and intra-cluster time excess delay, explained in \cite{Samimi16a}. Cluster powers are modeled as an exponentially decaying function of cluster time excess delays, and powers of MPCs in each time cluster are modeled as an exponentially decaying function of intra-cluster time excess delays. Note that the total received power is updated based on the time-variant large-scale path loss described in Section \ref{sec:lsp}.
\subsubsection{Smooth transitions}
LSPs and SSPs have been continuously updated in each channel segment, while the initial channel coefficients of channel segments are independently generated. Therefore, a smooth transition procedure is applied to ``connect'' channel segment by cluster birth and death. The power of one old cluster ramps down and one new cluster ramps up \cite{WinnerII}. Since the number of time clusters in two channel segments may be different, the power of one cluster can ramp up or down individually. Note that cluster birth and death only happen to one time cluster in a channel snapshot.

\subsection{Human Blockage Model}
A human blockage event usually causes a temporal shadowing loss, which may last about several hundreds of milliseconds. A typical blockage event can be divided into four stages, unshadowed, decay, shadowed, rising shown in Fig. 4 in \cite{Mac17b}, and a four-state Markov model corresponding to four stages was proposed to characterize blockage events \cite{Mac17b,Mac19a}. It is found from field human blockage measurements using three sets of antennas (7\degree, 15\degree, and 60\degree~half power beamwidth (HPBW)) \cite{Mac17b} that the transition rates in the Markov model and the mean attenuation of blockage events depend on antenna HPBW. More blockages and deeper blockages will occur if a narrower antenna is equipped at the receiver. Note that aforementioned human blockage measurements are on peer-to-peer level, and are not representative of a BS-UT setting. However, T-R separation distances are usually greater than 40 m, BSs are on the horizon and thus are at similar heights relative to the UT. Thus, blockage statistics abstracted from peer-to-peer measurements are reasonable to be extended to the BS-UT setting. A linear approximation for four transition rates with respect to the antenna HPBW is given as $\lambda_{\textup{decay}} = 0.2$, $\lambda_{\textup{shadow}} = 0.065\times\textup{HPBW}+7.425$, $\lambda_{\textup{rise}} = 0.05\times\textup{HPBW}+7.35$, and $\lambda_{\textup{unshadow}} = 6.7$. 

\begin{figure}[htbp]
	\centering
		\setlength{\abovecaptionskip}{-0.1cm}
		\setlength{\belowcaptionskip}{-0.7cm}
	\subfigure[CDFs of simulated human blockage shadowing loss using the four-state Markov model for omnidirectional channels in UMi LOS and NLOS scenarios.]{
		\begin{minipage}[b]{0.42\textwidth}
			\includegraphics[width=1\linewidth]{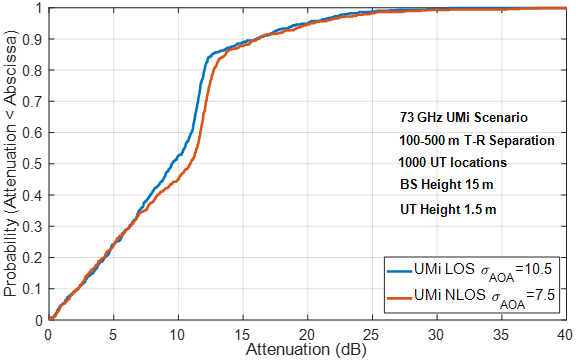}
			\label{fig:omni_bk}
						\vspace{-1.5em}
		\end{minipage}
	}
	\quad
	\subfigure[CDFs of simulated human blockage shadowing loss using the four-state Markov model for directional channels with RX antenna azimuth HPBWs (7\degree, 15\degree, 30\degree, and 60\degree)]{
		\begin{minipage}[b]{0.42\textwidth}
			\includegraphics[width=1\linewidth]{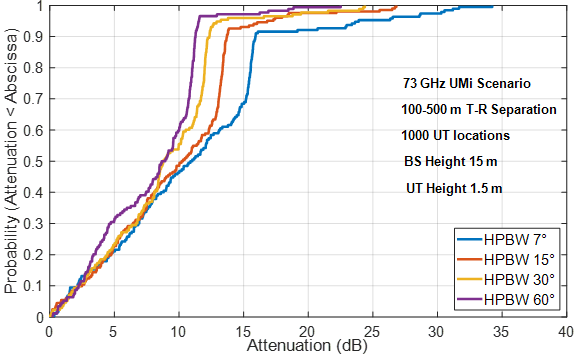}
			\label{fig:dir_bk}
						\vspace{-1.5em}
		\end{minipage}
	}
	\caption{CDFs of human blockage shadowing loss for omnidirectional and directional channels} 
	\label{fig:bk}
\end{figure}

However, for omnidirectional channels generated from NYUSIM, antenna HPBW is not applicable, spatial lobe width, instead, is used to calculate transition rates and mean attenuation. Spatial lobes represent main directions of arrival (or departure) there energy may arrive over several hundreds of nanoseconds. A spatial lobe is modeled as a Gaussian distribution, where the variance is lobe angular spread that varies according to scenario and environment. The azimuth lobe angular spread obtained from measurements is 10.5\degree~in a LOS scenario \cite{Samimi16a}. \textcolor{black}{Based on three-sigma rule of thumb, a typical width of a spatial lobe is less than 10.5\degree$\times$6 = 63\degree~with 99.7\% probability, which is comparable to the 60\degree~HPBW antenna used in the measurements \cite{Mac17b}. Thus, it is valid to assume that all MPCs in the same spatial lobe will experience approximately the same amount of shadowing loss by a blockage event.} In addition, a directional channel realization output from NYUSIM is generated by applying antenna patterns of both transmitter (TX) and receiver (RX) to the omnidirectional channel realization. The applied antenna pattern emulates the horn antenna pattern used in the mmWave field measurements \cite{Rap13a}, and can be calculated by the user-specified antenna azimuth and elevation HPBW. \textcolor{black}{For directional channels generated from NYUSIM, all MPCs within the antenna HPBW are assumed to experience the same blockage event and be attenuated by approximately the same amount.}

3GPP 38.901 proposed 5 potential independent blockers for each path \cite{3GPP.38.901}. NYUSIM assumes that $m$ independent blockage events (uniformly distributed between 1 and 5) may occur for each spatial lobe. $m$ independent blockage events are represented by $m$ independent Markov traces. $m$ Markov traces are superimposed to obtain a total loss trace. The actual shadowing loss is attained by randomly selecting a time instance $t_0$ on the total loss trace. Note that the aforementioned human blockage procedure is done independently for omnidirectional channel and directional channel in one simulation run, since the transition rates and average mean attenuation for two kinds of channels are different. A comparison of cumulative distribution functions (CDFs) of shadowing loss for omnidirectional channels in UMi LOS and NLOS scenarios is shown in Fig. \ref{fig:omni_bk}. The T-R separation distance range was from 100 m to 500 m. 1000 UT locations were simulated. The shadowing loss in the NLOS scenario is slightly larger than in the LOS scenario since the spatial lobe angular spread is smaller in the NLOS scenario, which means the average spatial lobe width is smaller in the NLOS scenario. In the NLOS scenario, the UT may experience more than 10 dB and 15 dB shadowing loss with 55\% and 12\% probabilities, respectively. A comparison of CDFs of shadowing loss for directional channels in the UMi NLOS scenario using four sets of antennas with HPBW 7\degree, 15\degree, 30\degree, and 60\degree~is shown in Fig. \ref{fig:dir_bk}. UTs equipped with narrower HPBW antennas are more likely to experience severe blockage shadowing loss. 31\% of UTs equipped with 7\degree~HPBW antenna experience more than 15 dB shadowing loss. To verify the simulated channel coefficients such as large-scale path loss and human blockage shadowing loss, the outage probability of generated directional channels is compared with the outage probability of measured directional channels. -5 dB SNR is considered as the outage threshold. SNR is given as $\textup{SNR}=\textup{P}_\textup{r}[\textup{dBm}]-(\textup{N}_0+\textup{NF})$, where $\textup{P}_\textup{r}$ is the total received power in dBm, which is equivalent to the sum of all MPCs' powers generated from NYUSIM. $\textup{N}_0$ is the average thermal noise power with 800 MHz RF bandwidth, which is -84.97 dBm, and noise figure (NF) is 10 dB \cite{Mac19a}. The comparison between simulations from NYUSIM and field measurements \cite{Mac19a} is given in Table. \ref{tab:sim_meas}. It is seen that the simulated outage probability and 5\% SNR level are close to the measured outage probability and 5\% SNR level, which indicates that the NYUSIM can accurately recreate realistic directional channels with human blockage shadowing loss.

\begin{table}
	\centering
	\caption{Simulated and measured CDF of user SNR values (in dB) with and without human blockage events at 73 GHz. 1000 simulations of directional channels were run in Monte Carlo fashion.}
	\label{tab:sim_meas}
	\begin{tabular}{|p{1.5cm}|p{1.3cm}|p{1.2cm}|p{1.3cm}|p{1.2cm}|}
\hline
& \multicolumn{2}{c|}{\textbf{Simulations}} & \multicolumn{2}{c|}{\textbf{Measurements \cite{Mac19a}}} \\ \hline
\textbf{Blockage?}   & \textbf{-5 dB SNR Thresh. \%}    & \textbf{5\% SNR}   & \textbf{-5 dB SNR Thresh. \%}              & \textbf{5\% SNR}             \\ \hline
\textbf{No blockage} & 14.7\%                           & -7.2 dB            & 16.5\%                                     & -10.8 dB                     \\ \hline
\textbf{Blockage}    & 25.5\%                           & -15.5 dB           & 24.7\%                                     & -17.7 dB                     \\ \hline
	\end{tabular}
\vspace{-0.5cm}
\end{table}

\subsection{O2I Penetration Loss Model}
A parabolic model for building penetration loss from \cite{A5GCM15,Haneda16a} is implemented in NYUSIM, which has either a low loss or a high loss form, depending on the type of building surface. The low loss model works for external building materials like standard glass and wood while the high loss model works for external building materials like IRR glass and concrete \cite{Haneda16a}. The parabolic model for building O2I penetration loss (BPL) is given as \cite{A5GCM15,Haneda16a}:
\begin{equation}
	\text{BPL[dB]}=10\log_{10}(A+B\cdot f_c^2)+N(0,\sigma_P^2)
\end{equation}
where $f_c$ is the carrier frequency. $A = 5$, $B = 0.03$, and $\sigma_P = 4.0$ for the low loss model.  $A = 10$, $B =5$, and $\sigma_P = 6.0$ for the high loss model.

\vspace{-0.1cm}
\section{NYUSIM 2.0}~\label{sec:gui}
\begin{figure*}[!htbp]
	\centering
		\setlength{\abovecaptionskip}{0.1cm}
	\setlength{\belowcaptionskip}{-0.6cm}
	\includegraphics[width=1\textwidth]{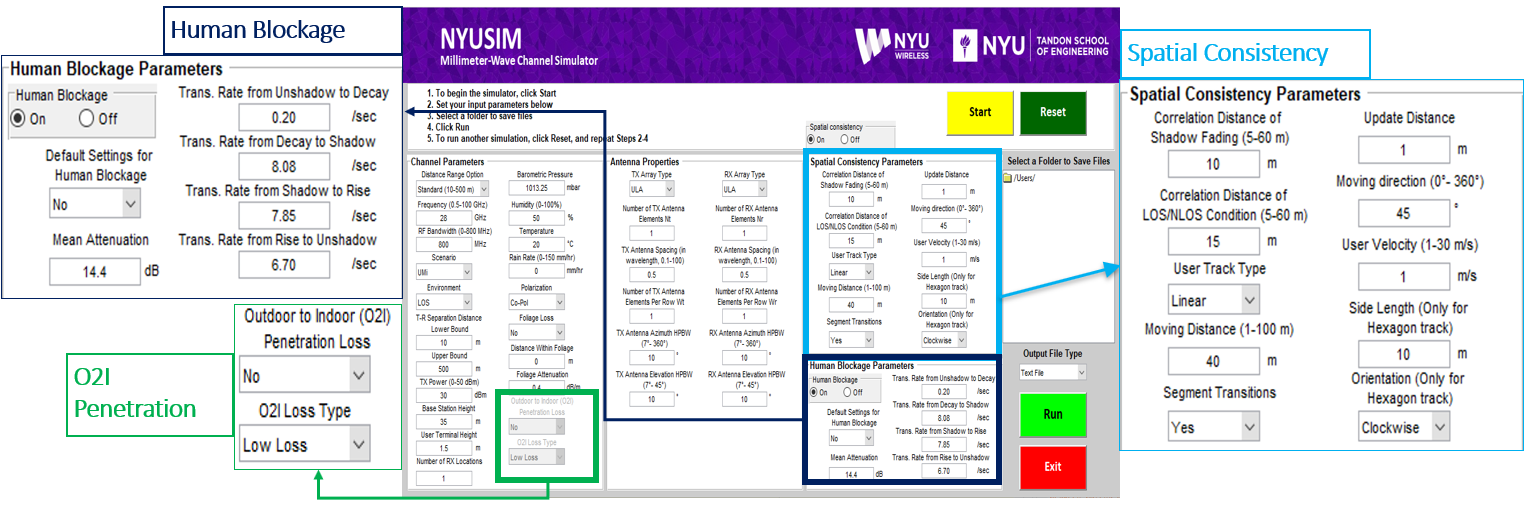}
	\caption{NYUSIM 2.0 GUI shows the three new dialogue boxes for spatial consistency, human blockage, and O2I penetration loss.} \label{fig:gui}
\end{figure*}

NYUSIM 2.0 with additional features such as spatial consistency, human blockage, and O2I penetration can generate continuous time-variant CIRs and simulate the power loss caused by random human blockage events and external building penetration. The screenshot in Fig. \ref{fig:gui} shows the graphical user interface (GUI) of NYUSIM 2.0. Two running modes, drop-based mode and spatial consistency mode are made available in NYUSIM 2.0. When the spatial consistency button is ``on'', NYUSIM runs spatial consistency procedure and generates successive and correlated CIRs along the UT trajectory. When the spatial consistency button is ``off'', NYUSIM runs the drop-based model which is the same as older versions of NYUSIM and generates independent CIRs for different distances. Note that human blockage module works for both drop-based mode and spatial consistency mode.

\subsection{Input parameters}
In addition to the existing 16 input ``Channel Parameters'' that define the propagation channel and 12 input ``Antenna Properties'' that specify the TX and RX antenna arrays, 10 new input ``Spatial Consistency'' parameters, five new input ``Human Blockage'' parameters, and two new input ``O2I penetration'' parameters are provided. Two types of user track, linear and hexagon, are provided for spatial consistency modeling. A ``Default Settings for Human blockage'' is provided. If the user chooses ``Yes'', the transition rates and average mean attenuation are implemented based on a linear fit to the data given in \cite{Mac17b}, and changes to fit the work in \cite{Mac17b} as the RX antenna azimuth HPBW changes. If the user chooses ``No'', the transition rates and mean attenuation are free to edit for user-specified preferences, in case other Markov model parameters are preferred or discovered in the future. 

\subsection{Sample Output Figures and Output Data}~\label{sec:output}
The default output figures for the drop-based mode are same as the figures generated in older versions of NYUSIM (e.g. 1.6.1 and earlier), as shown in \cite{Sun17b}. Some default output figures for the spatial consistency mode are listed below. A map of spatially correlated SF with the UT and BS locations and UT track is output as shown in Fig. \ref{fig:sf_map}. A map for spatially correlated LOS/NLOS condition using scenario-specific LOS probability is output and indicates that all locations in a local area experience the same propagation condition (LOS or NLOS). Consecutive omnidirectional and directional PDPs along the user trajectory are displayed to show the power variation and delay drifting of MPCs. Directional PDPs are obtained by applying antenna pattern in the strongest received MPC direction to the omnidirectional PDP. 3D AOA and AOD angular power spectrum is also provided to record the time-variant angles. All spatially correlated and time-variant channel parameters along the UT trajectory are stored in an output folder for further use such as beam steering algorithm test, dynamic channel condition, and MIMO system performance evaluation.
\begin{figure}[!htbp]
	\centering
	\setlength{\abovecaptionskip}{0cm}
	\setlength{\belowcaptionskip}{-0.4cm}
	\includegraphics[width=0.45\textwidth]{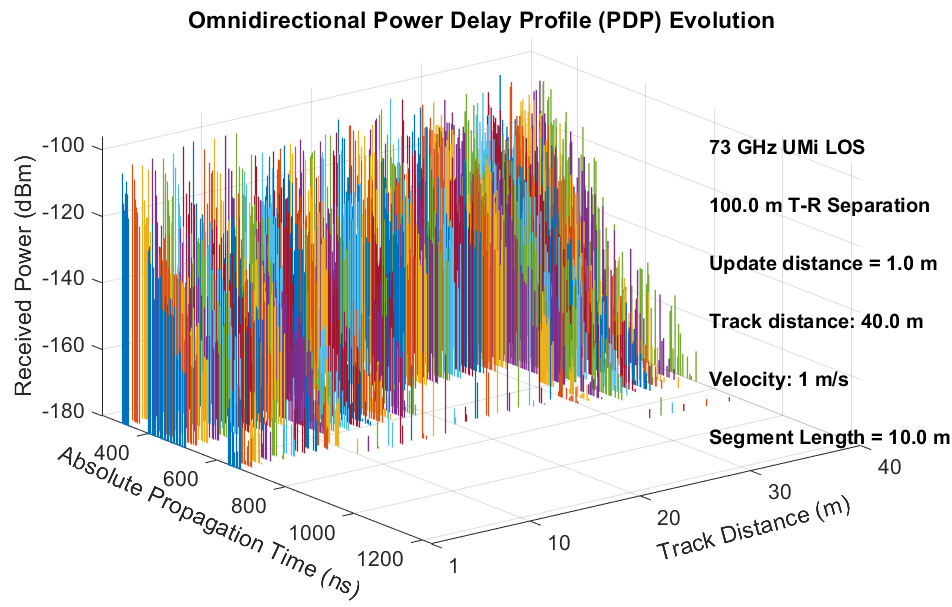}
	\caption{A typical output consecutive omnidirectional PDPs from NYUSIM 2.0. For each channel snapshot, a PDP with absolute time delay is stored and displayed.} \label{fig:conOmni_sam1}
\end{figure}
\section{Conclusion}\label{sec:conclusion}
An extended NYUSIM channel model with spatial consistency, human blockage and O2I penetration loss was developed. The multiple reflection surfaces method was adopted to update small-scale parameters in the spatial consistency procedure. A four-state Markov model was implemented to simulate human blockage events. Antenna HPBW for directional channels and spatial lobe width for omnidirectional channels were used to calculate transition rates and mean attenuation of blockage events. A parabolic model was implemented for O2I penetration loss, which will be useful for the future extended NYUSIM channel model for indoor scenarios. NYUSIM 2.0 can generate consecutive and correlated omnidirectional and directional PDPs based on the user-specified UT trajectory. 
\vspace{-0.2cm}
\bibliographystyle{IEEEtran}
\bibliography{gc}

\begin{thebibliography}{10}
\providecommand{\url}[1]{#1}
\csname url@samestyle\endcsname
\providecommand{\newblock}{\relax}
\providecommand{\bibinfo}[2]{#2}
\providecommand{\BIBentrySTDinterwordspacing}{\spaceskip=0pt\relax}
\providecommand{\BIBentryALTinterwordstretchfactor}{4}
\providecommand{\BIBentryALTinterwordspacing}{\spaceskip=\fontdimen2\font plus
\BIBentryALTinterwordstretchfactor\fontdimen3\font minus
  \fontdimen4\font\relax}
\providecommand{\BIBforeignlanguage}[2]{{%
\expandafter\ifx\csname l@#1\endcsname\relax
\typeout{** WARNING: IEEEtran.bst: No hyphenation pattern has been}%
\typeout{** loaded for the language `#1'. Using the pattern for}%
\typeout{** the default language instead.}%
\else
\language=\csname l@#1\endcsname
\fi
#2}}
\providecommand{\BIBdecl}{\relax}
\BIBdecl

\bibitem{Rap13a}
T.~S. Rappaport \emph{et~al.}, ``Millimeter wave mobile communications for {5G}
  cellular: It will work!'' \emph{IEEE Access}, vol.~1, pp. 335--349, May 2013.

\bibitem{Sun18a}
S.~Sun \emph{et~al.}, ``Propagation models and performance evaluation for {5G}
  millimeter-wave bands,'' \emph{IEEE Transactions on Vehicular Technology},
  vol.~67, no.~9, Sep. 2018.

\bibitem{Rap19access}
T.~S. {Rappaport} \emph{et~al.}, ``Wireless communications and applications
  above 100 {GHz}: Opportunities and challenges for {6G} and beyond,''
  \emph{IEEE Access}, vol.~7, pp. 78\,729--78\,757, Jun. 2019.

\bibitem{Ju18a}
S.~Ju and T.~S. Rappaport, ``Simulating motion - incorporating spatial
  consistency into the {NYUSIM} channel model,'' \emph{2018 IEEE 88th Vehicular
  Technology Conference Workshops}, pp. 1--6, Aug. 2018.

\bibitem{Ju18b}
------, ``Millimeter-wave extended {NYUSIM} channel model for spatial
  consistency,'' \emph{2018 IEEE Global Communications Conference (GLOBECOM)},
  pp. 1--6, Dec. 2018.

\bibitem{Wang15a}
Y.~{Wang} \emph{et~al.}, ``A millimeter wave channel model with variant angles
  under {3GPP SCM} framework,'' in \emph{IEEE 26th Annu. Inte. Symp. on
  Personal, Indoor, and Mobile Radio Comm.}, Aug 2015, pp. 2249--2254.

\bibitem{Ademaj17a}
F.~Ademaj, M.~K. Mueller, S.~Schwarz, and M.~Rupp, ``Modeling of spatially
  correlated geometry-based stochastic channels,'' in \emph{2017 IEEE 86th
  Vehicular Technology Conference (VTC-Fall)}, Sept 2017, pp. 1--6.

\bibitem{Mac17b}
G.~R. {MacCartney} \emph{et~al.}, ``Rapid fading due to human blockage in
  pedestrian crowds at {5G} millimeter-wave frequencies,'' in \emph{2017 IEEE
  Global Communications Conference}, Dec 2017, pp. 1--7.

\bibitem{Mac19a}
G.~R. MacCartney \emph{et~al.}, ``Millimeter-wave base station diversity for
  {5G} coordinated multipoint ({CoMP}) applications,'' \emph{IEEE Transactions
  on Wireless Communications}, May 2019.

\bibitem{Xing19a}
Y.~Xing \emph{et~al.}, ``Indoor wireless channel properties at millimeter wave
  and sub-{Terahertz} frequencies: Reflection, scattering, and path loss,'' in
  \emph{Proc. 2019 Global Communications Conferences}, Dec. 2019, pp. 1--6.

\bibitem{Haneda16a}
K.~{Haneda} \emph{et~al.}, ``{5G 3GPP-Like} channel models for outdoor urban
  microcellular and macrocellular environments,'' in \emph{2016 IEEE 83rd
  Vehicular Technology Conference (VTC Spring)}, May 2016, pp. 1--7.

\bibitem{Andrew17a}
J.~G. {Andrews} \emph{et~al.}, ``Modeling and analyzing millimeter wave
  cellular systems,'' \emph{IEEE Trans. on Comm.}, vol.~65, no.~1, pp.
  403--430, Jan 2017.

\bibitem{Sun17b}
S.~Sun \emph{et~al.}, ``A novel millimeter-wave channel simulator and
  applications for {5G} wireless communications,'' in \emph{2017 IEEE
  International Conference on Communications}, May 2017, pp. 1--7.

\bibitem{Rap15b}
T.~S. Rappaport \emph{et~al.}, ``Wideband millimeter-wave propagation
  measurements and channel models for future wireless communication system
  design ({Invited Paper}),'' \emph{IEEE Transactions on Communications},
  vol.~63, no.~9, pp. 3029--3056, Sept. 2015.

\bibitem{Samimi16a}
M.~K. Samimi and T.~S. Rappaport, ``{3-D} millimeter-wave statistical channel
  model for {5G} wireless system design,'' \emph{IEEE Trans. on Microwave
  Theory and Tech.}, vol.~64, no.~7, pp. 2207--2225, July 2016.

\bibitem{Sun16a}
S.~{Sun} \emph{et~al.}, ``Investigation of prediction accuracy, sensitivity,
  and parameter stability of large-scale propagation path loss models for {5G}
  wireless communications,'' \emph{IEEE Transactions on Vehicular Technology},
  vol.~65, no.~5, pp. 2843--2860, May 2016.

\bibitem{Mac15a}
G.~R. {Maccartney} \emph{et~al.}, ``Indoor office wideband mmwave propagation
  measurements and channel models at 28 and 73 {GHz} for ultra-dense {5G}
  wireless networks,'' \emph{IEEE Access}, vol.~3, pp. 2388--2424, 2015.

\bibitem{Liebe93a}
H.~Liebe \emph{et~al.}, ``Propagation modeling of moist air and suspended
  water/ice particles at frequencies below 1000 {GHz},'' in \emph{Atmospheric
  Propagation Effects Through Natural and Man-Made Obscurants for Visible to
  MM-Wave Radiation}, 1993.

\bibitem{3GPP.38.901}
3GPP, ``Technical specification group radio access network; study on channel
  model for frequencies from 0.5 to 100 {GHz (Release 15)},'' TR 38.901
  V15.0.0, June 2018.

\bibitem{3GPPTDOC1707267}
{3GPP}, ``Discussion on procedure {A} for spatially-consistency {UT} mobility
  modeling,'' ZTE, TDOC R1-1707267, May 2017.

\bibitem{Ju19b}
S.~Ju, ``Channel modeling and channel simulation for fifth-generation and
  beyond millimeter-wave wireless communications,'' \emph{New York University
  Library, Master of Science}, May 2019.

\bibitem{WinnerII}
P.~Ky{\"o}sti \emph{et~al.}, ``{WINNER} {II} channel models,'' European
  Commission, IST-WINNER, D1.1.2 V1.2, Feb. 2008.

\bibitem{A5GCM15}
5GCM, ``{5G} channel model for bands up to 100 {GHz},'' Oct. 21, 2016.

\end{thebibliography}

\end{document}